\documentclass[reprint,aps,pra,superscriptaddress,
amsmath,amssymb]{revtex4-1}
\usepackage{mathtools}
\usepackage{graphicx}
\usepackage{booktabs,multirow}
\providecommand{\JournalTitle}[1]{#1}
\usepackage{braket}
\usepackage{textcomp}
\graphicspath{ {figures_for_paper/} }
\usepackage[caption=false]{subfig}
\usepackage{braket}
\usepackage[
  bookmarks=false,
  colorlinks,
  linkcolor=blue,
  urlcolor=blue,
  citecolor=blue,
  plainpages=false,
  pdfpagelabels,
  final,
  breaklinks=true
]{hyperref}

\begin{document}

\preprint{APS/123-QED}
\title{Super-resolution with Fourier measurements}

\author{S. A. Wadood}
\affiliation{Department of Electrical and Computer Engineering, Princeton University, New Jersey 08544, USA}
\author{Shaurya Aarav}

\affiliation{Department of Electrical and Computer Engineering, Princeton University, New Jersey 08544, USA}
\author{Kevin Liang}
\affiliation{Physics Department, Adelphi University, Garden City, New York 11530, USA}

\author{Jason W Fleischer}
\email{jasonf@princeton.edu}
\affiliation{Department of Electrical and Computer Engineering, Princeton University, New Jersey 08544, USA}

\date{\today}

\begin{abstract}
Resolving sources beyond the diffraction limit is important in imaging, communications, and metrology. Current image-based methods of super-resolution require phase information (either of the source points or an added filter) and perfect alignment with the centroid of the object. Both inhibit the practical application of these methods, as uniform motion and/or relative jitter destroy their assumptions. Here, we show that measuring intensity in the Fourier plane enables super-resolution without any of the issues of image-based methods. We start with the shift-invariance of the Fourier transform and the observation that the two-point position problem ${x_1,x_2}$ in the near field corresponds to the single-point wavenumber problem $k\ =2\pi/(x_2-x_1)$ in the far field. We consider the full range of mutual coherence and show that for fully coherent sources, the Fourier method saturates the quantum limit, i.e. it gives the best possible measurement. Similar results hold for sub-Rayleigh constellations of $N$ sources, which can act collectively as a spatially averaged metasurface and/or individually as elements of a phased-array antenna.  The theory paves the way to merge Fourier optics with super-resolution techniques, enabling experimental devices that are both simpler and more robust than previous designs.  
\end{abstract}

\maketitle
\begin{figure}[ht!]
\centering
\hspace*{-0.75cm} 
\includegraphics[trim={0cm 0cm 0cm 0cm},clip,scale=0.47]{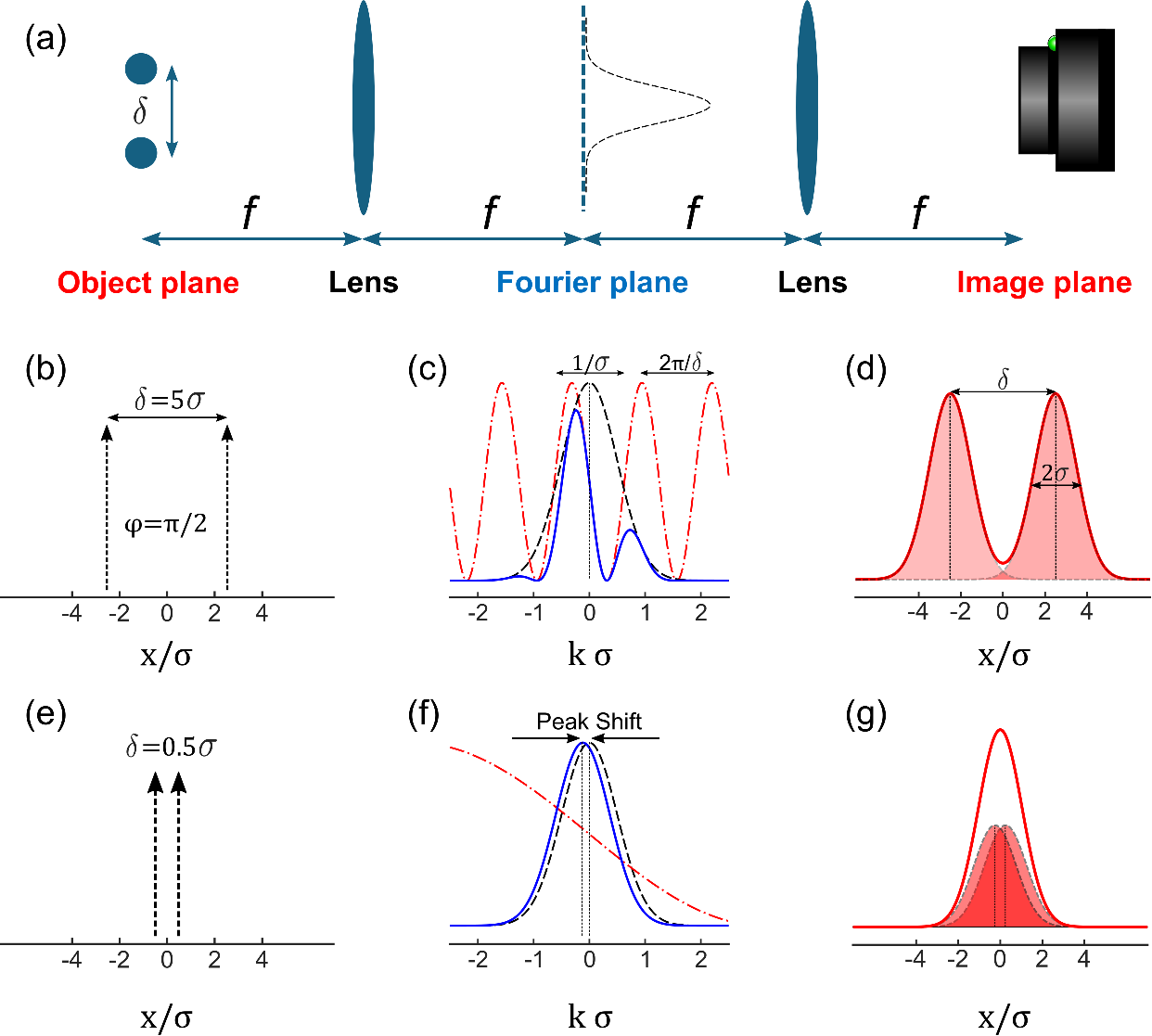}
\caption{Basic idea of Fourier super-resolution. (a) Two point sources with separation $\delta$ imaged by a $4-f$ imaging system with a Gaussian aperture in the Fourier plane, and a spatially resolving CCD in the image plane. (b) Imaging two point sources with $\delta=5\sigma$ showing well-resolved direct imaging in the image plane (d). (e) For sub-Rayleigh $\delta=0.5\sigma$, the two-point intensity in the image plane (g) is sub-resolving, and estimating $\delta$ with DI is sub-optimal. The two point intensity in (d, g) is denoted by the solid red line, while the individual wavepacket intensities are denoted by dashed black lines. The vertical dashed lines in the image plane denote the peak location $\pm\delta/2$ for the two Gaussians. In this paper, we propose measuring intensity in Fourier space (c, f). The Fourier intensity in (c) shows the $k-$space oscillations (dashed-dotted red line) due to two-point interference. (f) shows the Fourier intensity for the the sub-Rayleigh regime of (g). While the $k-$space oscillations are much broader than the Gaussian pupil function, they still cause a finite shift of the spectrum peak from the origin in $k-$space. Localizing this peak can provide sub-Rayleigh resolution. For this figure, $\gamma=1$ and the relative phase $\phi=\pi/2$.\if The $k-$space shift varies with $\phi$, and the inset shows the spectrum centroid versus $\phi/\pi$ for $\delta=0.5\sigma$.  Note that for all phases except $\phi=0,\pi$, the centroid is shifted from $k=0$.\fi}
\label{fig:1}
\end{figure}

\section{Introduction}
Discrimination of signals overlapping in space or time is fundamental to optical imaging and communications. However, all practical optical systems have finite apertures, which restricts the amount of information available for detection. In the ideal limit of point sources, the bounded transfer function (known as the point spread function, or PSF) limits the ability to isolate a single point or discriminate between two closely spaced ones. The corresponding resolution criteria, due to Abbe \cite{abbe1873articlebeitrage} and Rayleigh \cite{rayleigh1879xxxi}, respectively, have guided the design of optical systems for more than a century.

While the simplest approach to improve resolution is to increase the aperture size, this is not always practical nor does it solve the fundamental wave problem from physical optics. Instead, current super-resolution schemes exploit other degrees of freedom, e.g. source details  \cite{hellSTED1994breaking,pavani2009doublehelix,rust2006subSTORM}, near-field capture \cite{betzig1991breakingnearfieldmicroscopy,sanchez1999nearfieldnovotny,he2021ultrahigh}, nonlinearity \cite{Jason2013nonlinear_Abbe_theory}, and optical \cite{tsang2016quantumtheoryofsuperresolution,GburReview} or  digital \cite{dertinger2009SOFI,hugodefienne2022pixelsuperresolution} processing. All these techniques rely on nontrivial engineering of the object or image-plane field. Here, we provide a simple alternative that does not require any change to a canonical imaging setup. Specifically, we show that measuring intensity in the Fourier plane enables super-resolution of two or more separated points.\par

The results follow by realizing that two separated points uniquely define a wavevector, with the magnitude given by their separation distance and the direction defined by the line joining them. This converts the two-point problem $\{x_1,x_2\}$ in image space into a single-wavenumber problem ${k\ =2\pi/(x}_2-x_1)$ in Fourier space. Working in the Fourier domain also gives several advantages, including 1) a direct description in terms of waves, 2) a phase sensitivity arising from the preferred $k$, and 3) a shift-invariance that makes detection robust to the misalignment issues that compromise image-based methods.

Below, we discuss the fundamentals of our approach. In Section 2, we present our representative geometry and give a brief description of super-resolution efforts in image space. In Section 3, we describe our new Fourier method for two-point discrimination, its performance in the presence of noise, and its generalization to constellations of multiple points. In Sections 4 and 5, we discuss applications and avenues for future work.

\section{Image-plane super-resolution}

The geometry of a canonical imaging system is shown in Fig. (1a). For simplicity, we consider one transverse dimension, a $4-f$ imaging system with unit magnification, and a Gaussian PSF that maps a point source to a symmetric blur of width $\sigma$. Accordingly, two point sources separated by $\delta$ in the object plane map to two Gaussian PSFs in the image plane, still separated by a distance $\delta$ (Fig. (1b)).  In the super-Rayleigh case of $\delta>\sigma$, the two Gaussians are well resolved by direct imaging (DI) of the image-plane intensity (Fig. (1d)). In the sub-Rayleigh regime of $\delta<\sigma$ (Fig. (1e)), however, the image-plane intensity of the two Gaussians is not well resolved (Fig. (1g)), and, in the presence of noise, DI cannot afford super-resolution.\if To first order in $\delta$, the intensity profile of the two $\delta$-separated PSFs does not change and, in the presence of noise, DI cannot afford super-resolution. \par \fi
\par Intuitively, one would expect super-resolution by designing a measurement scheme whose response (PSF)varies appreciably in the limit $\delta\ll\sigma$. To this end, progress can be made only if there is more information about the system. The most restrictive case, and the one used by all early attempts at super-resolution, assumes two point sources of equal amplitude and a known phase relation, with an imaging system that is aligned along the centerline (centroid) of the object. This wealth of information allows one to reduce the task from a functional discrimination of PSFs to an estimation of a single parameter $\delta$. For example, if it is known that the two point sources are coherent with a relative phase $\Delta\phi=\ \pi$, then there is always an intensity null between the image points, and $\delta$ can be estimated by photon counting \cite{wadood2021experimentalparitysorting,giacomokaruseichyk2022resolvingmutuallycoherentpointsources}. If the object points are mutually incoherent, then the introduction of a $\pi$ phase mask (e.g. a knife-edge filter) and a single-mode detector can accomplish the same \cite{tsang2016quantumtheoryofsuperresolution,steinberg2017beating_rayleighscurse}. Over time, the necessary conditions on the object scene have been relaxed \cite{vrehavcek_sanchez_soto_2017multiparameter,kevin2021partialcoherencemomentsPRA}, but the need for system alignment has remained. Unfortunately, this severely limits the practical application of image-based methods, as any uniform motion or relative jitter upsets the required symmetry.

\section{Fourier-plane super-resolution}

While Fourier optics lies at the foundation of single-point resolution, \'a la Abbe \cite{goodman_Fourier_Optics_book}, its application to the two-point problem has received scant attention. It is surprising that a switch from the image to Fourier plane would give any advantage, as a change in basis should not alter the fundamentals of the problem; Fourier transforms are unitary, and modes in position ($x$-space) and momentum ($k$-space) are equivalent representations. Indeed, it is an elementary problem in physical optics to show that Rayleigh’s definition of resolution in direct imaging can be derived from Abbe’s approach using the angular spectrum.\par 

The emphasis on localization, however, breaks the symmetry between the representations. In particular, a narrow distribution in the image plane corresponds to a broad and flat one in the Fourier plane. More rigorously, the shift invariance of the Fourier transform means that the $k$-space basis is inappropriate for identifying the source position.\par

Remarkably, this intuition does not hold when two point sources are considered. In this case, the sources give rise to a Fourier spectrum that is determined uniquely by their separation distance and \textit{relative} orientation. Using the Fourier relation between the PSF and pupil \textit{fields} in the paraxial approximation \cite{goodman_Fourier_Optics_book}, we find that the Fourier \textit{intensity} for two point sources transforms to a single Gaussian shifted in proportion to $\delta$. Spectral measurements therefore convert two-point discrimination to single-point localization tasks. Notably, single-point localization is unconstrained by the Rayleigh limit and is more cost-effective (from an optoelectronic perspective) than two-point localization. 

In Section A, we present a fundamental derivation of two-point super-resolution from a Fourier perspective. In Section B, we use information theory to quantify the performance of the method in the presence of noise. Finally, in Section C, we  generalize the approach to constellations of N source points. 

\subsection{Derivation}

\if While the quantitative measure and advantage of super-resolution differs for each modality and application, the overarching goal is to devise a measurement that is sensitive to changes in the object scene parameters.  Consider the two-point imaging problem, where the goal is to estimate the separation $\delta$ between the two sources.  
\par Intuitively, one would expect super-resolution with any measurement scheme whose response versus $\delta$ varies appreciably in the limit $\delta\ll\sigma$. To first order in $\delta$, the intensity profile of the two $\delta$-separated PSFs does not change and, in the presence of noise, DI cannot afford super-resolution. Instead of looking at the position space intensity, we propose monitoring the Fourier space intensity, i.e., the spectrum.\par 
The result is surprising, as a simple switch in perspective should not alter the fundamentals of the problem. Indeed, Fourier transforms are unitary, and modes in position ($x$-space) and momentum ($k$-space) are equivalent representations. It is an elementary problem in physical optics to show that Rayleigh’s definition of resolution in direct imaging can be derived from Abbe’s approach using the angular spectrum.\par

Both the resolution limit and the equivalence between representations can be broken if more information is added to the system. For example, if it is known that the object is a single point source, then the imaging problem can be rephrased as the task of finding the position of a delta function. As no imaging system is perfect, the delta function is blurred to a PSF. It is then useful in practice to interpret the resulting PSF as a probability distribution for photodetection and convert the task to one of estimating its centroid. This is much less demanding than determining the exact shape of the PSF *(esp. when it is circularly symmetric ALSO HINTS OF 2D)*.\par
The emphasis on localization also breaks the symmetry between the $x$-space and $k$-space representations. In particular, a narrow distribution in the image plane corresponds to a broad and flat one in the Fourier plane. More rigorously, the shift invariance of the Fourier transform means that the latter basis is inappropriate for identifying the source position.\par

Remarkably, this intuition does not hold when two point sources are considered. In this case, the sources give rise to a Fourier spectrum that is determined uniquely by their separation distance and relative orientation (HINTS AT 2D SO SHOULD WE DISCUSS). While the PSF and pupil \textit{fields} are related by a unitary Fourier transform in the paraxial approximation \cite{goodman_Fourier_Optics_book}, we find that the Fourier \textit{intensity} for two coherent point sources in the sub-Rayleigh regime can transform to a single Gaussian shifted in proportion to $\delta$. Spectral measurements therefore convert sub-Rayleigh two-point localization to single point localization tasks. Notably, single point localization is unconstrained by the Rayleigh limit and is more cost-effective, from an optoelectronic perspective, than two-point localization. We now consider a Fourier measurement of the two-point problem in detail.\par\fi

Consider two Gaussian PSFs 
\begin{align}
f_{\pm}=E_{\pm}(2\pi\sigma^2)^{-1/4}\exp[-(x\pm\delta/2)^2/4\sigma^2]
\end{align}
\noindent separated in space by $\delta$, where $E_{\pm}$ are the complex field amplitudes and $x$ represents the spatial coordinate. The Fourier transform is given by
\begin{align}
\tilde{F}_{\pm}(k)=\frac{E_{\pm}}{\sqrt{2\pi}}\int dx e^{ikx}f_{\pm}(x)=E_{\pm}(2\sigma^2/\pi)^{1/4}e^{\mp\frac{ik\delta}{2}-k^2\sigma^2}, 
\end{align}
\noindent where $k$ is the corresponding spatial frequency. Assuming equal emission of $M/2$ photons and a fixed relative phase $\phi$ between the sources, the Fourier-plane intensity $I(k,\delta)=\langle|\tilde{F}_{+}(k)+\tilde{F}_{-}(k)|^2\rangle$ reads 
\begin{align}
    I(k,\delta)=M(2\sigma^2/\pi)^{1/2}e^{-2k^2\sigma^2}\left(1+\gamma\cos{(k\delta+\phi)}\right)\label{eq:IntensityFT},
\end{align}
\noindent where the ensemble-averaged cross-correlation ${\langle E_{+}E^{*}_{-}}\rangle=\frac{M}{2} \gamma e^{i\phi}$ has been expressed in terms of the mutual coherence (visibility) $0\leq\gamma\leq1$.\par

Equation (\ref{eq:IntensityFT}) is a fundamental, well-known result of physical optics. The Gaussian envelope represents the pupil function, and the cosine factor is the far-field interference pattern of two point sources. From an information theory perspective, the envelope characterizes the optical system, while the interference encodes details of the object. Indeed, if the sources are mutually incoherent $(\gamma = 0)$, then information about their separation does not appear in the Fourier intensity. To illustrate the interference effect more clearly, we consider here the "best case" scenario of fully coherent sources (partial coherence is treated in Appendix A). 

For $\gamma=1$, Eq. (\ref{eq:IntensityFT}) expresses a competition between the separation distance $\delta$ and the width $\sigma$ of the PSF. In the super-Rayleigh case, $\delta>\sigma$, Eq. (\ref{eq:IntensityFT}) predicts spectral oscillations with a period smaller than the Gaussian pupil function (Figs. (\ref{fig:1}c)). In the sub-Rayleigh case, the period of spectral oscillations is much broader than the pupil width (Figs. (\ref{fig:1}e-g)). As we are interested in super-resolved measurements, we restrict our attention to the latter case.

For a given separation distance, the peak of the Gaussian intensity depends on the relative phase $\phi$ between the sources. For example, a relative phase $~\phi=\pi/2$ gives the Fourier-plane intensity 
\begin{align}
    I(k)&\propto e^{-2k^2\sigma^2}\left(1-k\delta\right)\approx e^{-2\sigma^2(k+k_{max})^2},\label{eq:shift_FT_intensity}
\end{align}
\noindent where the wavenumber displacement $k_{max}=\frac{\delta}{4\sigma^2}$ is assumed to be $\ll1/4\sigma$. 

The original problem of determining the separation $\delta$ between two $x$-space (image-plane) Gaussians with unknown centroid is now reformulated as a problem of determining the shift of a \textit{single} $k$-space Gaussian (Fig. (\ref{fig:1}f)). Direct imaging of the Fourier intensity (with a continuum of pixels) then provides an optimal estimation of the spectral centroid \cite{tsang2016quantumtheoryofsuperresolution}. In our 1D example,  even a 2-pixel split detector can obtain state-of-the-art performance for localizing the Gaussian spectrum in the presence of shot noise \cite{Trepshsu2004optimaldisplacementofsingleGaussian}.\par 

\begin{figure}[ht]
\centering
\hspace*{-1cm} 
\includegraphics[trim={0cm 0cm 0cm 0cm},clip,scale=1]{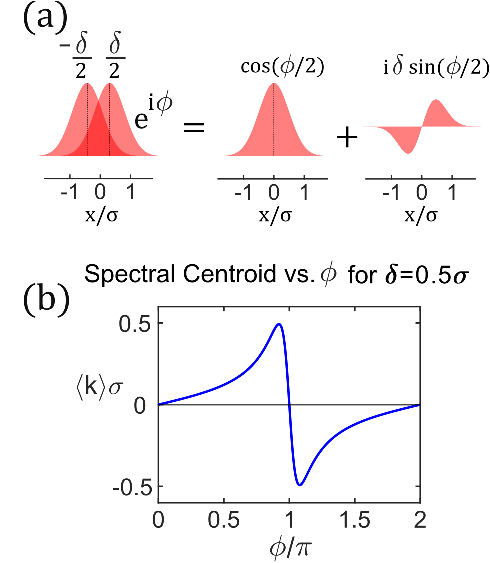}
\caption{(a) Modal decomposition of shifted Gaussians. For $\delta/\sigma\ll1$, a shifted Gaussian can be represented as sum of an on-axis Gaussian and a first order Hermite--Gauss mode whose amplitude is proportional to the shift. \if For the coherent superposition of two Gaussians separated by $\delta$ with a phase $\phi$, the on-axis, even mode has an amplitude proportional to $\cos(\phi/2)$ and the first order, odd mode has an amplitude proportional to  $i\delta\sin(\phi/2)$. \fi Because the two modes are in quadrature, the perturbative first order mode does not contribute to the intensity, at least to first order in $\delta$. In the Fourier domain, due to the Gouy phase shift, the modes are in phase and the spectrum is shifted. \if For $\phi=\{0,\pi\}$, there is no spectral shift because the mode is either fully symmetric ($\phi=0$) or antisymmetric ($\phi=\pi$).\fi The $k-$space shift varies with $\phi$, and (b) shows the spectral centroid versus $\phi/\pi$ for $\delta=0.5\sigma$. Note that for all phases except $\phi=0,\pi$, the centroid is shifted from $k=0$. For this figure, $\gamma=1$.} 
\label{fig:2}
\end{figure}

To get more insight into the $\delta-$dependent shift, we now compare the modal content in $x$-space with its conjugate spectrum in $k$-space. In the image plane, a Taylor expansion of the field in $\delta$ yields a superposition of an on-axis, zeroth-order Gaussian with amplitude proportional to $\cos(\phi/2)$ and a first-order mode with amplitude proportional to $i\delta\sin(\phi/2)$ (Fig. (\ref{fig:2}a)) \cite{tsang2016quantumtheoryofsuperresolution}. Because the first-order mode is in quadrature with the zeroth-order mode, it does not contribute to the image-plane intensity (if terms of $O(\delta^2)$ are ignored). In the Fourier domain, however, the first-order mode picks up an extra $\pi/2$ phase and becomes in-phase with the zeroth-order Fourier mode, resulting in a shifted spectrum. (Note that Hermite--Gauss modes retain their shape upon Fourier transformations, and for Gaussian PSFs the $\pi/2$ phase gained by the first-order mode could be attributed to the Gouy phase). Equivalently, the $\pi/2$ phase is due to the Fourier relation replacing $\partial_{x}\xrightarrow[]{}-ik$ for arbitrary, well-behaved PSFs. Its action is reminiscent of the $\pi/2$ focal plane phase mask in Zernike's phase-contrast microscopy \cite{goodman_Fourier_Optics_book}. 

From Eq. (\ref{eq:IntensityFT}), it can be shown that the $k$-space centroid of the normalized Fourier intensity is equal to $\braket{k}=\beta k_{max}$, where $\beta=d\sin(\phi)/\left(1+d\cos\left(\phi\right)\right)$ for $d=\exp{[-\delta^2/8\sigma^2]}$. The Fourier-plane intensity therefore shifts for all phases $\phi\neq\{0,\pi\}$ and thus can resolve sub-Rayleigh separations. (For $\phi=\{0,\pi\}$, the Fourier-plane intensity does not shift, as there is a maximum (symmetric mode $\phi=0$) or minimum (antisymmetric mode $\phi=\pi$) at $k=0$ for all $\delta$.) An example of spectral centroid shift of $\braket{k}$ for $\delta=0.5\sigma$ is plotted in Fig. (\ref{fig:2}b). \par 
\if
For partially coherent sources with $\gamma<1$, the effects discussed above would be less pronounced, with ultimately no advantage for the incoherent case of $\gamma=0$ for which the intensity in Eq. (\ref{eq:IntensityFT}) does not depend on $\delta$. We focus on the completely coherent case for the following sections. \par
\fi

\begin{figure}[ht!]
\centering
\hspace*{-1cm} 
\includegraphics[trim={0cm 0cm 0cm 0cm},clip,scale=0.65]{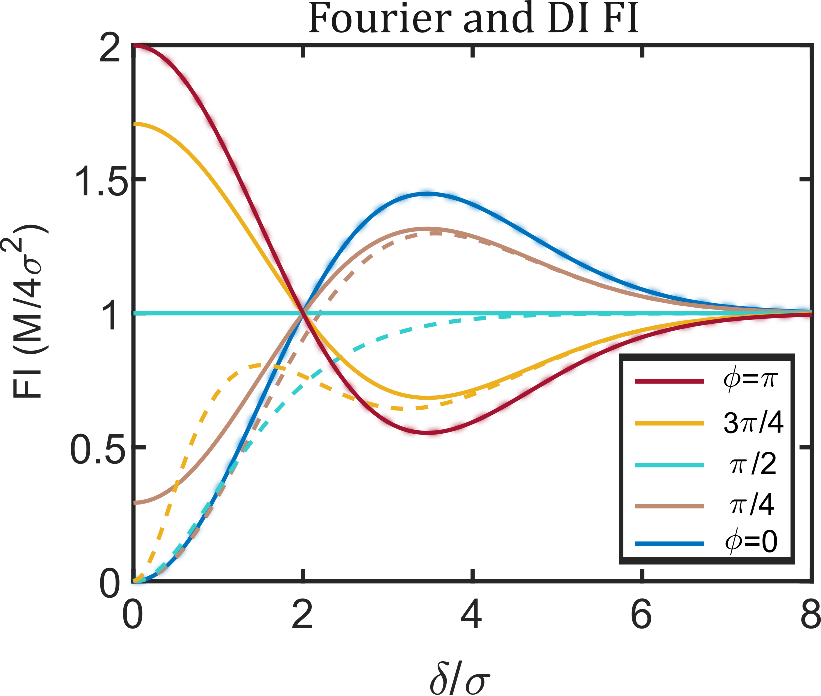}
\caption{ FI (Eq. (\ref{eq:FIversusdelta})) for the Fourier plane measurement in units of $M/4\sigma^2$ and for different $\phi$ values. Solid lines indicate the FI for Fourier intensity measurement, and dashed lines indicate the FI for DI. Non-zero FI values for $\phi\neq0$ in the limit $\delta\to0$ show evasion of Rayleigh's curse. The Fourier curves retain the same structure for any misalignment of the centroid from the optical axis. For  $\phi=\{0,\pi\}$, the DI curves coincide with the solid Fourier curves, and are slightly blurred for distinguishability.}
\label{fig:3}
\end{figure}

\subsection{Fisher Information}
The physical optics arguments made above for estimating $\delta$ do not include the effect of noise. The conclusions are therefore insufficient, as the SNR is infinite in the absence of noise, and there is no problem of super-resolution. To rigorously quantify resolution, we use the information-theoretic approach of \cite{tsang2016quantumtheoryofsuperresolution}. We are interested in a precise estimation of the unknown parameter $\delta$, assuming all other parameters $M,\sigma,\phi$ are known. Specifically, the variance in any (unbiased) estimate of $\delta$ is bounded from below by the Cramer-Rao inequality \begin{align}
    Var\{\hat{\delta}\}\geq F\left(\delta\right)^{-1},\label{eq:CRB}
\end{align} where $\hat{\delta}$ is an estimate of the true value $\delta$, and $F(\delta)$ is the (single parameter) Fisher information (FI) associated with the particular measurement scheme \cite{anthonyvella_alonso_2020MLE_fisher_information_tutorial}. Eq. (\ref{eq:CRB}) shows that an optical measurement with a higher FI can provide more precise estimation of $\delta$ than a scheme with lower FI. For diffraction-limited direct imaging (DI), the FI goes to zero, or the error in an (unbiased) estimate blows up, as $\delta\rightarrow0$. Often called Rayleigh's curse \cite{tsang2016quantumtheoryofsuperresolution} this behavior is an information-theoretic rendition of Rayleigh's criterion. Any optical technique claiming to achieve super-resolution needs to have a non-zero FI in the sub-Rayleigh regime $\delta\ll\sigma$. 

The FI for estimating $\delta$ from the Fourier intensity measurement scheme, henceforth called Fourier FI, can be found from the relation 

\begin{align}
F(\delta,\phi) =   \int dk\left(\partial_{\delta}I\left(k,\delta\right)\right)^2/I\left(k,\delta\right), 
\end{align}

\noindent where $I(k,\delta)$ is given by Eq. (\ref{eq:IntensityFT}) and governs the underlying probability distribution of photons impinging on the Fourier plane \cite{anthonyvella_alonso_2020MLE_fisher_information_tutorial}, and we have assumed shot noise-limited Poisson statistics for intensity \cite{tsang2016quantumtheoryofsuperresolution}. We obtain the closed-form expression (for $\gamma = 1$)
\begin{align}
F(\delta,\phi) 
&=\frac{M}{4\sigma^2}\left(1+\cos{(\phi)}e^{-\delta^2/8\sigma^2}(\delta^2/4\sigma^2-1)\right)\label{eq:FIversusdelta}.
\end{align}
\normalsize
\noindent In the sub-Rayleigh limit of $\delta\xrightarrow[]{}0$, the per-photon FI is
\begin{align}
        \frac{1}{M}\lim_{\delta\to0}F(\delta,\phi)=\frac{1}{4\sigma^2}\left(1-\cos{\left(\phi\right)}\right),\label{eq:FI_coherent}
\end{align}
\noindent which is non-zero for $\phi\neq0$. To the best of our knowledge, this is the first exposition of two-point super-resolution in an unmodified (unfiltered) coherent imaging system. \if a classical scheme that evades Rayleigh's curse for coherent sources \cite{Caveat_DI_anticorrelated}. \fi Figure (\ref{fig:3}) shows plots of Eq. (\ref{eq:FIversusdelta}) for sources that are positively correlated ($\phi=\{0,\pi/4\}$), anticorrelated ($\phi=\{3\pi/4,\pi\}$) and in quadrature ($\phi=\pi/2$). \par

Figure (\ref{fig:3}) also shows the FI for DI with coherent sources, calculated using the image-plane intensity $I(x,\delta)=|\langle E_{+}(x)+E_{-}(x)(x)\rangle|^2$. For PSFs with no nulls, such as the Gaussian PSF considered here, the FI for DI vanishes quadratically with $\delta$ for $\delta\xrightarrow[]{}0$ for all $\phi\neq\pi$. (For PSFs with zeros, the nulls lead to linear scaling \cite{Sanchez_Soto_Tempering_Rayleighs_Curse_2018Optica}, but the FI still vanishes for $\delta\xrightarrow[]{}0$ for all $\phi\neq\pi$). The anticorrelated case ($\phi=\pi$) is more subtle. Strictly speaking, DI with anticorrelated sources can also evade Rayleigh's curse, but this advantage stems from the total photon number measurement, i.e., spatially resolved detection is not required for that case. This is because the field for $\phi=\pi$ is always in an antisymmetric mode (Fig. \ref{fig:2}a); the integrated intensity scales as $1-\exp[-\delta^2/8\sigma^2]$, and both DI and Fourier measurements can provide nonvanishing FI in the sub-Rayleigh case by integrating over their respective intensities (for more details on anticorrelated DI, see \cite[Figs.(4,6),]{giacomokaruseichyk2022resolvingmutuallycoherentpointsources} and \cite{wadood2021experimentalparitysorting,kurdzialek2022backtosourcesroleofloss}). The Fourier advantage is most evident for complex coherence, i.e.,  $\{\phi\neq0,\pi\}$. In this case, Fourier measurements always beat DI and offer nonzero FI in the sub-Rayleigh regime. \if Only the Fourier method, however, can accommodate alignment jitter or phase fluctuations.\fi 

\subsection{$N$ coherent sources}
\if The Fourier advantage easily generalizes to a constellation of $N$ coherent sources, a case encountered in many diverse applications. These range from microwave phased antenna arrays \cite{schelkunoff1943mathematicaltheoryoflineararrays}, where the goal is to optimize the far-field radiation pattern, to X-ray crystallography \cite{bornandwolf2013principlesofoptics}, which can be modelled as structured radiation from arrays of atoms in a crystal, to metasurface design \cite{nikolov2021metaformnickvamivakasmetaformdesign}, where optical antennas are engineered to achieve exotic wavefront control, and to coupled atomic arrays and sensors probing many-body physics \cite{mandelandwolf1995book,bernien2017probingLukin51atomquantumsimulator,masson2020manyOrozcomanybodysignaturesinatomicchains,changKimble2012cavityQEDwithatomicmirrors,asenjo2019optical,rovnynatalie2024nanoscalenvcentersreviewarticle}. \fi \if These include microwave phased antenna arrays \cite{schelkunoff1943mathematicaltheoryoflineararrays}, X-ray crystallography \cite{bornandwolf2013principlesofoptics}, metasurface design \cite{nikolov2021metaformnickvamivakasmetaformdesign}, and coupled atomic arrays and sensors probing many-body physics \cite{mandelandwolf1995book,bernien2017probingLukin51atomquantumsimulator,masson2020manyOrozcomanybodysignaturesinatomicchains,changKimble2012cavityQEDwithatomicmirrors,asenjo2019optical,rovnynatalie2024nanoscalenvcentersreviewarticle}. More precisely, we can ask if the Fourier plane measurement is useful for measuring the effective width $L$ of a constellation of $N$ coherent sources in the sub-Rayleigh regime. We use a simplified model below to answer this question.\fi \if Does the same qualitative behavior for two sources,  where anticorrelation provides more information on the separation,  also hold for the N source case? .\fi\par

\begin{figure}[!]
\centering
\hspace*{-0.25cm} 
\includegraphics[trim={0cm 0cm 0cm 0cm},clip,scale=0.35]{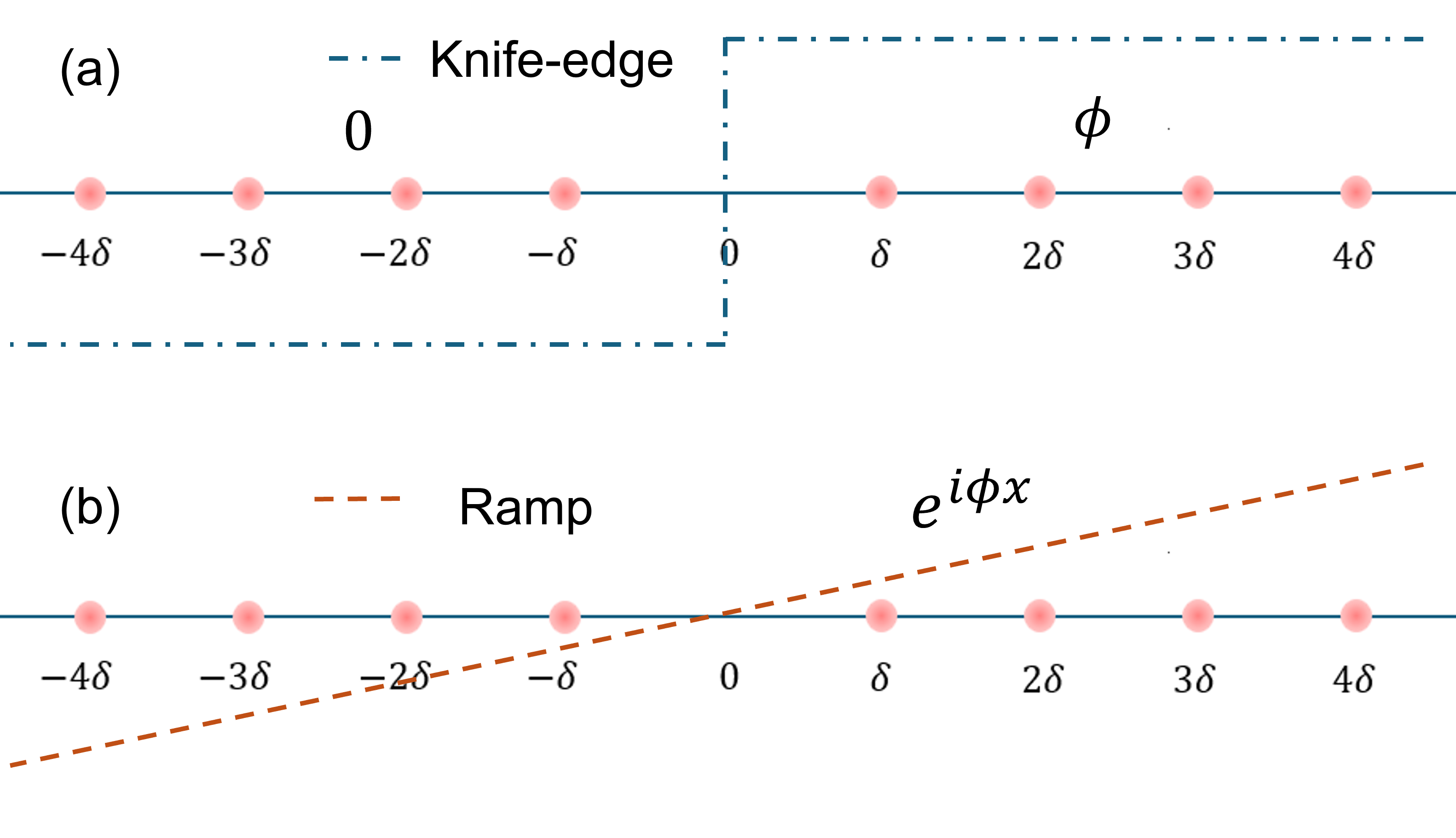}
\caption{Model constellation of $N=8$ point sources. The distance between each source on one side is $\delta$. (a) For a knife-edge style phase structure shown in dashed-dotted blue line, the sources on the right each have a phase of $\phi$ relative to the sources on the left. (b) For a ramp phase, the phase accrues linearly across the constellation given by $\exp{[i\phi x]}$. Note that there is no source at the origin.} 
\label{fig:4}
\end{figure}

The Fourier advantage easily generalizes to a constellation of $N$ coherent sources. To start, we consider an array of $N$ point sources, $N$/2 on the left and $N$/2 on the right side of the origin,  where the extent of the object $L=N \delta$ is the unknown parameter to be estimated (Fig. (\ref{fig:4})). For a knife-edge style phase, each source on the right has a phase of $\phi$ relative to the sources on the left.  We state here the final result of our analysis and defer the mathematical details to Appendix B. In the sub-Rayleigh regime and in the limit of large $N$, the per-photon FI to estimate $L$ is $NF_{N=2}/8$, where $F_{N=2}$ is the two-point FI given by Eq. (\ref{eq:FI_coherent}). \textit{The FI increases linearly with the number of sources}. This linear relation is not restricted to a knife-edge phase, but also holds for a tilt phase, i.e., a linear phase ramp. For both the knife-edge and ramp phase, the FI is maximized when the opposite edges of the constellation are anticorrelated. For an arbitrary wavefront impinging on the constellation in the sub-Rayleigh regime, we can expect the odd-ordered components of the wavefront to locally act as a tilt/phase ramp and contribute a non-zero FI. The even order components, because they do not naturally shift the Fourier spectrum, would be expected to have negligible contribution to the FI. \par

\section{Discussion}
The FI for Fourier measurements remains the same for arbitrarily high levels of misalignment. This is not just an information-theoretic bonus, but also i) reduces the prior knowledge required for estimation and ii) significantly relaxes the alignment constraints, making estimation immune to both static and dynamic misalignment in the object scene. \par
In addition to being robust to centroid fluctuations, spectral measurement is not complicated to implement. \if It is not complicated to implement in both spatial and temporal optical systems.\fi In the paraxial approximation, far-field propagation yields a Fourier transform. Furthermore, a lens naturally performs the spatial Fourier transform \cite{goodman_Fourier_Optics_book}. The results apply equally to the temporal domain where the temporal separation between two coherent pulses can be estimated by measuring the spectrum of the pulse train. \if[MOVE TO SPADE DISCUSSION]This is arguably less complicated than performing SPADE in time domain, which might require nonlinear interactions \cite{donohue2018quantumlimitedtimefrequencyestimation,SanchezSoto_Silberhorn2021effectsofcoherenceontemporalresolution} for mode projections.\fi\par
 As such, the Fourier measurement is a specific example of super-resolution via interferometric imaging, utilizing the phase of the coherent field. The Fourier plane intensity is a propagation-induced far-field interference pattern of the $N$-point constellation \cite{goodman2015statisticaloptics}. While a complete interferometric measurement is optimal for arbitrary incoherent imaging \cite{PRL_lupo2020quantumlimitstoincoherentimagingachievedbyinterferometry}, coherence in the sources obviates the need for a stable reference, beam splitting, or measurement of the cross-spectral density. Coherence, or phase, is therefore essential to link amplitude in the image plane to intensity in the Fourier plane, a fact underpinning most of (coherent) Fourier optics \cite{goodman_Fourier_Optics_book}. As mentioned earlier, partial coherence ($\gamma<1$) or phase fluctuations reduce the visibility of the cosine in Eq. (\ref{eq:IntensityFT}). The Fourier advantage is therefore expected to diminish with incoherence, lowering the Fourier FI as is shown in Appendix A. Nevertheless,
we show that Fourier measurements can outperform DI
as long as the sources are not completely incoherent, and the
correlation phase $\phi\neq \{0,\pi\}$. \par
Our analysis can be readily extended to more complex scenes. Although we have focused on the knife-edge and ramp phase relation between the sources, it is entirely plausible that a different combination of phases further increases the FI \cite{Kevinliang2023quantum_N_point_partially_coherent_sources}. This amounts to optimizing the (structured) illumination in coherent optical systems. Applications range from microwave phased antenna arrays \cite{schelkunoff1943mathematicaltheoryoflineararrays}, where the goal is to optimize the far-field radiation pattern, to X-ray crystallography \cite{bornandwolf2013principlesofoptics}, which can be modelled as structured radiation from arrays of atoms in a crystal, to metasurface design \cite{nikolov2021metaformnickvamivakasmetaformdesign}, where optical antennas are engineered to achieve exotic wavefront control, and to coupled atomic arrays and sensors probing many-body physics \cite{mandelandwolf1995book,bernien2017probingLukin51atomquantumsimulator,masson2020manyOrozcomanybodysignaturesinatomicchains,changKimble2012cavityQEDwithatomicmirrors,asenjo2019optical,rovnynatalie2024nanoscalenvcentersreviewarticle}. 

Another extension is to include multiple unknown parameters beyond the separation $\delta$ as the only unknown parameter. We can include cases of sources with unknown strengths,  varying distances between sources, arbitrary coherence structures, and extend to 2D and 3D imaging and classification \cite{michaelgrace2021quantumoptimal_obect_classification,zhang2024performance_partiallycoherenthypothesistesting,zhang2025quantum_partiallycoherenthypothesistesting}. Progress in the multiparameter estimation problem will, in turn, inform the important task of sub-Rayleigh phase retrieval \cite{fienup1982phaseretrieval,he2021ultrahigh}.\par 

 While we have analyzed the scalar regime, vector generalization is straightforward. The spectral shift there would be qualitatively similar to weak measurements, where postselection on a particular polarization state can result in an anomalously large shift in the pointer wavefunction \cite{dresselandrewjordan2014colloquiumweakvalues}.\if Of course, our analysis self-consistently takes into account the low probability of weak value amplification. \fi \par 

Finally, we note that the FI for the methods discussed in this work depend on their respective measurement schemes and are also referred to as classical FI. An important question to ask, then, is how much the classical FI can be improved. Optimized for all possible measurements, the upper bound on the classical FI, also called the quantum FI, informs us on the optimality of a particular scheme. If a classical FI saturates the quantum FI, the scheme is optimal, and further improvements are not possible on the detection end \cite{tsang2016quantumtheoryofsuperresolution,sethlloydgiovannetti2011advancesinquantummetrology}. An important feature of Fourier measurements is that they saturate the quantum FI for coherent sources \cite{kurdzialek2022backtosourcesroleofloss,tsang2021poisson,giacomokaruseichyk2022resolvingmutuallycoherentpointsources}. Fourier measurements are, by design, phase-sensitive and utilize the complete information in the image-plane field. Because of their tolerance to misalignment, Fourier measurements are more efficient than other well-established phase-sensitive techniques that also achieve sub-Rayleigh resolution, e.g. those using demultiplexing, sorting, or projection of image-plane spatial modes \cite{tsang2016quantumtheoryofsuperresolution}. In Appendix C, we give derivations for these latter
approaches in a form that allows direct comparison between
them and the Fourier approach. Like DI, however, the advantage of these image-plane methods is susceptible to noise, cross-talk, and misalignment \cite{tsang2016quantumtheoryofsuperresolution,Michael_Grace_JOSAB_unknown_Centroid,vrehavcek_sanchez_soto_2017multiparameter,hervas2024optimizingsanchezsotoetradeoffsinmultiparametersuperresolution,de2021discrimination_under_misalignment,schodt2023tolerance_to_aberration_and_misalignment,optimal_observables_for_practical_superresolution,PRL2020_superresolution_limits_frm_crosstalk}. Fourier measurements therefore offer a robust alternative to achieve quantum-limited coherent super-resolution.

\if Typically implemented by projecting the image plane field to an appropriate modal basis \cite{OptExpress_Parity_Sorting2016,wadood2021experimentalparitysorting,Fabre2020_Optica_SPADE_2D_Crosstalk}, modal projections allow both spatial \cite{tsang2016quantumtheoryofsuperresolution,Steinberg2017BeatingRayleighscurse,superresolution_with_heterodyne,OptExpress_Parity_Sorting2016,sanchezsoto_2016_optica} and temporal \cite{SanchezSoto_Silberhorn2021effectsofcoherenceontemporalresolution} super-resolution of incoherent and coherent sources \cite{SalehResurgencePaper,tsangcommentonSaleh,SalehreplytoTsang,Kevin2021coherenceOptica,wadood2021experimentalparitysorting,hradil2021exploringtheultimatelimits,SanchezSoto_fisherinformation_With_Coherence_Optica,kurdzialek2022backtosourcesroleofloss,SanchezSoto_Silberhorn2021effectsofcoherenceontemporalresolution,giacomokaruseichyk2022resolvingmutuallycoherentpointsources,giacomosorelli2025ultimate}, and for arbitrary objects in the sub-Rayleigh regime \cite{zhou2019modern,tsang2019quantum,datta2021sub_rayleigh_characterization,sorelli2021momentbasedsuperresolution,frank2023passivelvovskypassivesuperresolution,Lvovsky_pushkina2021superresolution_PRL,Yiyu2021confocal_mode_sorter,tsang2020semiparametric,tsang2019semiparametric}. Although robust to partial or complete incoherence \cite{tsang2016quantumtheoryofsuperresolution,tsang2021poisson,wadood2021experimentalparitysorting,SalehreplytoTsang}, processing of spatial modes is susceptible to noise, cross-talk, and misalignment \cite{Michael_Grace_JOSAB_unknown_Centroid,vrehavcek_sanchez_soto_2017multiparameter,hervas2024optimizingsanchezsotoetradeoffsinmultiparametersuperresolution,de2021discrimination_under_misalignment,schodt2023tolerance_to_aberration_and_misalignment,optimal_observables_for_practical_super-resolution,PRL2020_super-resolution_limits_frm_crosstalk}. Specifically, modal projection is sub-optimal if the centroid of the object scene is unknown or misaligned, and Rayleigh's curse resurges for even a finite misalignment of the optical setup from the object centroid \cite{tsang2016quantumtheoryofsuperresolution,Michael_Grace_JOSAB_unknown_Centroid,vrehavcek_sanchez_soto_2017multiparameter,hervas2024optimizingsanchezsotoetradeoffsinmultiparametersuperresolution}.  \fi 

\if We show in Appendix C that the per-photon FI for an aligned mode sorter \if in the limit $\delta\xrightarrow[]{}0$\fi \textit{coincides exactly with the Fourier FI} for the two-point \cite{tsangcommentonSaleh,wadood2021experimentalparitysorting,Kevin2021coherenceOptica,kurdzialek2022backtosourcesroleofloss} and $N$-point case. Fourier measurements therefore offer a robust alternative to saturate the quantum FI for coherent super-resolution \cite{tsang2021poisson,giacomokaruseichyk2022resolvingmutuallycoherentpointsources,kurdzialek2022backtosourcesroleofloss}.\fi

\if Our work establishes that Fourier measurements are also optimal for coherent super-resolution, offering the same FI as quantum-inspired techniques under similar constraints.\fi \par 
\if
While it known that SPADE saturates the upperbound on FI, also called the quantum FI \cite{tsang2021poisson}, our works shows that Fourier measurements are also optimal for coherent super-resolution, offering the same FI as SPADE under the same constraints. In practice, either scheme might be preferable depending on the application and details of the object scene. Although SPADE is robust to partial or complete incoherence \cite{tsang2016quantumtheoryofsuperresolution}, its advantage is susceptible to noise, cross-talk, and misalignment \cite{Michael_Grace_JOSAB_unknown_Centroid,vrehavcek_sanchez_soto_2017multiparameter,hervas2024optimizingsanchezsotoetradeoffsinmultiparametersuperresolution,de2021discrimination_under_misalignment,schodt2023tolerance_to_aberration_and_misalignment,optimal_observables_for_practical_super-resolution,PRL2020_super-resolution_limits_frm_crosstalk}. SPADE is sub-optimal if the centroid of the object scene is unknown or misaligned, and Rayleigh's curse resurges for even a finite misalignment of the SPADE setup from the object centroid \cite{tsang2016quantumtheoryofsuperresolution,Michael_Grace_JOSAB_unknown_Centroid,vrehavcek_sanchez_soto_2017multiparameter,hervas2024optimizingsanchezsotoetradeoffsinmultiparametersuperresolution}. Specifically, if the object centroid is displaced from the optical axis by $\xi$, the FI drops to zero for $\delta\xrightarrow[]{}0$ for nonzero $\xi$, as shown in Fig. (\ref{fig:2}) for SPADE with HG modes up to order $20$ \cite{Fabre2020_Optica_SPADE_2D_Crosstalk}. While Fig. (\ref{fig:2}) is plotted with $\xi=0.1\sigma$, higher level of misalignment will further decrease the FI (solid lines) compared to the perfectly aligned SPADE (dashed lines). As with Fourier measurements, anticorrelation still yields a higher SPADE FI than correlated sources \cite{Caveat_DI_anticorrelated}.\par
\fi

\section{Conclusion}
We have shown that the Fourier basis is optimal for super-resolution with coherent sources. \if Under similar constraints, the Fourier measurement has the same FI as SPADE to estimate the separation between N symmetric, coherent point sources. \fi A Fourier measurement does not require knowledge of the object scene centroid, and is immune to centroid jitters. \if It is an interesting duality that while SPADE is robust to phase fluctuations, Fourier measurements are robust to centroid fluctuations.\fi Whether a measurement exists that tolerates both phase and centroid fluctuations is an open question. For general imaging tasks, we expect obtaining a robust improvement over DI using the full phase space as compared to working in either the position or momentum basis \cite{laurawaller2012phasespacemeasurementandcoherencesynthesis}.\par

\if Although more work needs to be done to see if the benefits of Fourier measurement also hold in the above-mentioned realistic scenarios,\fi Our work strongly suggests immediate benefits. Fourier measurements 1) are robust to static and dynamic jitters in the object scene, 2) generalizable to complicated objects, 3) utilize phase information, 4) are inherently multimodal in spatial and temporal applications, and 5) saturate the ultimate bounds set by quantum mechanics, making them uniquely equipped to provide robust improvements to spatio-temporal imaging and metrology. \if For general imaging under low SNR, not only does the measurement basis need to be optimal in the information-theoretic sense, it is also desirable to have minimal prior knowledge and overhead resources to perform the estimation. This is an open problem, but we can expect improvement over DI using the full phase space as compared to using either the position or momentum basis \cite{laurawaller2012phasespacemeasurementandcoherencesynthesis}.\fi
\subsection*{Acknowledgements}
SAW acknowledges M. A. Alonso and A. N. Vamivakas for useful discussions.
\subsection*{Funding}
 This work was supported by the Air Force Office of Scientific Research grant FA9550-23-1-0221, and National Science Foundation Program 25-535: Launching Early-Career Academic Pathways in the Mathematical and Physical Sciences, Award Number 2532835.
\small
\renewcommand\theequation{A\arabic{equation}}
\setcounter{equation}{0}
\subsection*{Appendix A: Fourier measurements with partial coherence}

We begin with Eq.(3) from the main text:

\begin{align}
    I(k,\delta)&=\langle|\tilde{F}_{+}(k)+\tilde{F}_{-}(k)|^2\rangle,\nonumber\\
    &=\left(\frac{2\sigma^2}{\pi}\right)^{1/2}e^{-2k^2\sigma^2}\left(\langle |E_{+}|^2\rangle+\langle |E_{-}|^2\rangle+2\Re\{\langle E_{+}E^{*}_{-}\rangle e^{-ik\delta}\}\right)\nonumber\\
    &=M\left(\frac{2\sigma^2}{\pi}\right)^{1/2}e^{-2k^2\sigma^2}\left(1+\gamma\cos{(k\delta+\phi)}\right)\label{eq:sup:IntensityFT},
\end{align}
where $\langle |E_{\pm}|^2\rangle=M/2$, i.e., each source emits $M/2$ photons, ${\langle E_{+}E^{*}_{-}}\rangle=\Gamma \sqrt{\langle|E_{+}|^2\rangle\langle|E_{-}|^2\rangle}=\frac{M}{2}\gamma e^{i\phi}$, and the angle brackets denote ensemble averaging. \if Note that Eq. (\ref{eq:IntensityFT}) of the main text follows from Eq. (\ref{eq:sup:IntensityFT}) for the completely coherent case of $\gamma=1$.\fi \par
We can now calculate the FI for estimating $\delta$, assuming all other parameters $M,\sigma, \gamma,\phi$ are known and assuming Poisson statistics for intensity \cite{tsang2016quantumtheoryofsuperresolution}. We have
\begin{align}
    \partial_{\delta}I=-M\left(\frac{2\sigma^2}{\pi}\right)^{1/2}e^{-2k^2\sigma^2}k\gamma\sin{(k\delta+\phi)},
\end{align}
so that the FI is given by

\begin{align}
F(\delta,\gamma,\phi)&=\int dk\frac{(\partial_{\delta}I)^2}{I}\nonumber\\&=M\left(\frac{2\sigma^2}{\pi}\right)^{1/2} \int dk e^{-2k^2\sigma^2}k^2\gamma^2\frac{\sin^2{(k\delta+\phi)}}{1+\gamma\cos{(k\delta+\phi)}}\label{eq:supplement_2_pt_FI}
\end{align}
For $\gamma=1$, the expression has the closed form solution given in Eq. (\ref{eq:FIversusdelta}) in the main text. For the case of low coherence, i.e., $\gamma\ll1$, the denominator in Eq. (\ref{eq:supplement_2_pt_FI}) can be approximated as unity, and we obtain 
\begin{align}
 F(\delta,\gamma\ll1,\phi)=\frac{M\gamma^2}{8\sigma^2}\left(1+\cos{(2\phi)}e^{-\delta^2/2\sigma^2}(\delta^2/\sigma^2-1)\right). \label{eq:sup:low_gamma_FI}  
\end{align}
Equation (\ref{eq:sup:low_gamma_FI}) shows that the Fourier FI scales as $\gamma^2$ for highly incoherent sources. Thus the Fourier FI is optimal for coherent sources, and has vanishing information for incoherent sources. In the limit $\delta\xrightarrow[]{}0$, Eq. (\ref{eq:supplement_2_pt_FI}) can be evaluated in closed form to give
\begin{align}
\lim_{\delta\to0}F(\delta,\gamma,\phi)&=M\left(\frac{2\sigma^2}{\pi}\right)^{1/2}\frac{\gamma^2\sin^2{(\phi)}}{1+\gamma\cos{(\phi)}} \int dk e^{-2k^2\sigma^2}k^2,\nonumber\\
    &=\frac{M}{4\sigma^2}\frac{\gamma^2\sin^2{(\phi)}}{1+\gamma\cos{(\phi)}}\label{eq:sup:FI_partialcoherent}.
\end{align}
\if Eq. (\ref{eq:sup:FI_partialcoherent}) is plotted in Fig. (\ref{fig:5}).\fi The FI is nonzero for partially coherent light ($\gamma\neq0$) with $\phi\neq 0$ (not completely in-phase). Rayleigh's curse is avoided for all points not on the $\textrm{Im}(\Gamma)=0$ line, and anticorrelation affords a higher FI than correlation. Note that the FI is zero on the real line ($\textrm{Im}(\Gamma)=0$) except for the case of $\gamma=1,\phi=\pi$. \par 
For the general case of $0\leq\gamma\leq1$, although a closed form expression for Eq. (\ref{eq:supplement_2_pt_FI}) does not exist. However, we can achieve a series solution by performing a cosine Fourier expansion of the even, periodic factor of  $\sin^2{(a)}/\left(1+\gamma\cos{\left(a\right)}\right )$,  where $a=k\delta+\phi$. We obtain

\begin{align}
    \frac{\sin^2{(a)}}{1+\gamma\cos{\left(a\right)}}&=\sum_{m=0}^{\infty}c_{m}\cos{(ma)},\label{eq:Fourier_series_integrand}\\
    c_{m}&=\frac{4}{1-\gamma}\left( _3\tilde{F}_2\left[1,\frac{3}{2},2;2-m,2+m;\frac{2\gamma}{\gamma-1}\right]\right.\nonumber\\
    & \left. -3~_3\tilde{F}_2\left[1,\frac{5}{2},3;3-m,3+m;\frac{2\gamma}{\gamma-1}\right] \right),
\end{align}
where $_3\tilde{F}_{2}[a_1,a_2,a_3;b_1,b_2;z]$ is a regularized hypergeometric function. Inserting Eq. (\ref{eq:Fourier_series_integrand}) in Eq. (\ref{eq:supplement_2_pt_FI}), the integral for each $m^{th}$ order cosine harmonic can be evaluated analytically and the series summed over $m$.

\par
\if
The FI for phase-sensitive schemes such as spatial mode demultiplexing (SPADE), in contrast to Eq. (\ref{eq:sup:FI_partialcoherent}), is given in the limit $ \delta\to0$ as \cite{tsangcommentonSaleh,Kevin2021coherenceOptica}
\begin{align}
\lim_{\delta\to0}    F_{PS}(\delta,\gamma,\phi) = \frac{M}{4\sigma^2}\left(1-\gamma \cos{(\phi)}\right)\label{eq:QFI},
\end{align}
which, per-photon, coincides with Eq. (\ref{eq:FI_coherent}) for $\gamma=1$, and is plotted in Fig. (\ref{fig:6}), and the subscript $PS$ denotes phase-sensitive. Unlike the Fourier FI, the FI for mode sorting schemes depends only on the real part of $\Gamma$. We plot the ratio of Eq. (\ref{eq:sup:FI_partialcoherent}) and Eq. (\ref{eq:QFI}) in Fig. (\ref{fig:7}), and see that at the edges of the disk ($\gamma=1$), the Fourier FI equals the mode sorting FI, i.e. the Fourier plane measurement provides the same advantage as phase-sensitive schemes for completely coherent sources. Note that for incoherent sources ($\gamma=0$), and on the real line where $\gamma\sin(\phi)=0$, the Fourier FI is zero and phase-sensitive detection on the image plane is still the optimal measurement. However, we stress that any finite coherence (with $\gamma>0, \phi>0$) can afford a nonzero Fourier FI \textit{for all $\delta$}, as shown by Eqs. (\ref{eq:sup:low_gamma_FI},\ref{eq:sup:FI_partialcoherent}).
\fi
\renewcommand\theequation{B\arabic{equation}}
\setcounter{equation}{0}
\subsection*{Appendix B: Fourier measurement of $N$ coherent point sources}
In this section, we calculate the FI for $N$ coherent point sources in the sub-Rayleigh regime. We consider an array of $N$ point sources, $N/2$ on the left and $N/2$ on the right side of the origin, as shown in Fig. (\ref{fig:3}).\if The simplifying assumptions for this model are described in the caption of Fig. (\ref{fig:3}), but we repeat them here for completeness.\fi The number of sources $N$ is assumed to be even, and the distance between each source on one side is $\delta$. Each source emits $M$ photons. 

We consider first a knife-edge phase, i.e. a configuration in which each source on the right has a relative phase of $\phi$ with each source on the left, while all sources on one side are in-phase with each other. We neglect any diffraction effects from the phase step; alternatively, the illumination optics are not assumed to be band-limited. For reasons of symmetry with respect to the phase, we have not included a source at the origin. The conclusions do not change if we place a point source with a prescribed phase between $0$ and $\phi$ at the origin, especially in the limit of large $N$ (the distinction between even and odd number of sources vanishes in this limit). Note that the maximum distance between two of the point sources is $N\delta=L$, where $L$ is the extent of the object. Both $\delta$, and hence $L$, are the only unknown parameters to be estimated. In the temporal domain the problem maps to a train of $N$ pulses, where the pulses are separated by a delay much smaller than the pulse width. \par
Each point source is assumed to map to a Gaussian PSF of width $\sigma$, shifted to the same location as the point source. The field at the object plane can then be written as
\begin{align}
    E(x) &=\sum_{j=-N/2}^{-1}A_{j}e^{-\frac{(x-j\delta)^2}{4\sigma^2}}+\sum_{j=1}^{N/2}A_{j}e^{-\frac{(x-j\delta)^2}{4\sigma^2}}e^{i\phi_{j}}\label{eq:N_point_field}
\end{align}
where $A_{j},\phi_{j},\delta_{j}$ are the amplitude, phase and location of the $j^{th}$ source, respectively. For simplicity, we assume that the strength of each source is equal, i.e. $A_{j}=A$ for all $j$. The Fourier transform of the object-plane field is then 
\begin{align}
     \tilde{E}(k)&=\sqrt{2}\sigma A e^{-k^2\sigma^2}\left(\sum_{j=1}^{N/2}e^{-ikj\delta }+e^{i\phi}\sum_{j=1}^{N/2}e^{ikj\delta }\right)\nonumber\\
    &=2\sqrt{2}\sigma A e^{-k^2\sigma^2}e^{i\phi/2}\frac{\sin{\left(\frac{kN}{4}\delta\right)}}{\sin{\left(\frac{k}{2}\delta\right)}}\cos{\left(\left(N+2\right)\frac{k\delta}{4}+\frac{\phi}{2}\right)},\label{eq:N_sources_Fourier_field}
\end{align}
where we have used the geometric sum $\sum_{j=0}^{N/2}e^{ikj\delta }=\frac{1-e^{ik\delta}e^{ikN\delta/2}}{1-e^{ik\delta}}$. The Fourier-plane intensity $\tilde{I}(k)$ is given by
\begin{align}
   \tilde{I}(k)=|\tilde{E}(k)|^2=8\sigma^2|A|^2e^{-2k^2\sigma^2}\frac{\sin^2{\left(\frac{kN}{4}\delta\right)}}{\sin^2{\left(\frac{k}{2}\delta\right)}}\cos^2{\left(\left(N+2\right)\frac{k\delta}{4}+\frac{\phi}{2}\right)}.\label{eq:Fourier_plane_intensity_N_sources}
\end{align}
In the limit $\delta \rightarrow 0$, the FI is
\begin{align}
       \lim_{\delta\to0}F(\delta)=M\frac{N^2(N+2)^2}{16\sigma^2}\frac{\left(1-\cos{(\phi)}\right)}{2},
\end{align}
where we have swapped the limit $\delta\to0$ with the integral over $k$ and set $|A|^2=M/\sqrt{2\pi}\sigma$. The FI to estimate $L=N\delta$ is $1/N^2$ times the FI for estimating $\delta$, since $\frac{d}{dL}\rightarrow\frac{d}{Nd\delta}$. The FI to estimate $L$ is then
\begin{align}
      \lim_{L\to0}  F(L)=  \lim_{\delta\to0}  \frac{F(\delta)}{N^2}=M\frac{(N+2)^2}{16\sigma^2}\frac{\left(1-\cos{(\phi)}\right)}{2}.\label{eq:FI_coherent_N_sources_delta}
\end{align}
The FI is non-zero for $\phi\neq0$. This shows that a structured phase relation in an extended (discrete) object allows one to estimate the length of the object in the sub-Rayleigh regime. As expected, anti-correlation provides better FI than positive correlation.\par
The per-photon FI for estimating $L$ is found by dividing Eq. (\ref{eq:FI_coherent_N_sources_delta}) by the total emitted photon number $NM$:
\begin{align}
    \lim_{L\to0}\frac{ F(L)}{NM} =\frac{(N+2)^2}{N}\frac{\left(1-\cos{(\phi)}\right)}{32\sigma^2}.\label{eq:FI_coherent_N_sources}
\end{align}
For $N=2$, $L$ is the separation between two sources, and we obtain the two-point result $F_{N=2}$ in Eq. (\ref{eq:FI_coherent}). For $N\gg2$, Eq. (\ref{eq:FI_coherent_N_sources}) reduces to $NF_{N=2}/8$,, and we see that the FI increases linearly with the number of sources. 

In this derivation, we have assumed that the only unknown parameter is the separation between the sources, and we knew both the number of sources and the number of \textit{emitted} photons. These are restrictive constraints, but our aim was to show the benefit of the Fourier measurement over DI with the same constraints. In Appendix C, we show that phase-sensitive (Hermite--Gauss) projections also give the same FI as Eq. (\ref{eq:FI_coherent_N_sources}).\if ([Comment 1: We have assumed N to be known. If N is unknown, then we also need to calculate the multiparameter FI. That case is more relevant for measuring the second moment of an extended object. Preliminary calculations show that even if N is unknown, the FI matrix is diagonal in the limit $\delta\rightarrow 0$, and so the inverted Cramer-Rao bound is non-zero, and is in fact equal to Eq. (\ref{eq:FI_coherent_N_sources}). Should this calculation be included, or not, perhaps in the supplement? Comment 2: We have also not normalized the total power, i.e., each source adds $M$ photons. We can add a prefactor of $1/\sqrt{M}$ to Eq. (\ref{eq:N_point_field}). The multiparameter FI still does not change from Eq. (\ref{eq:FI_coherent_N_sources}). Comment 3: Are there connections/benefits to viewing the constellation in terms of image charges where the phase is related to retardation?])\fi
\par

The intuition behind the non-zero FI obtained for the $N$-point case is similar to the two-point case, viz. the problem is transformed from measuring the separation $\delta$ between Gaussians in the image plane to measuring the shift (or centroid) of a single Gaussian in the Fourier plane. This can be seen easily by considering the sub-Rayleigh ($k_{max}\ll1/4\sigma$) Fourier-plane intensity in Eq. (\ref{eq:Fourier_plane_intensity_N_sources}) in the small $\delta$ limit:
\begin{align}
    \tilde{I}(k)=8\sigma^2|A|^2e^{-2k^2\sigma^2}\frac{N^2}{4}\left(\cos^2{(\phi/2)}-\sin{(\phi)}\frac{(N+2)k\delta}{4}\right),
\end{align}
which is a Gaussian whose peak is shifted from the origin in $k$ to $k_{max}=(1+\frac{2}{N})\tan{(\phi/2)}\frac{L}{8\sigma^2}$. For large $N$, $k_{max}=\tan{(\phi/2)}\frac{L}{8\sigma^2}$. For $\phi\neq\pi$, the shift of the Gaussian intensity profile in the Fourier plane is directly proportional to $L$.  \par

We now consider a ramp phase, where the $j^{th}$ source on the right (left) has a phase of $e^{ ij\phi} (e^{-ij\phi})$. In this case, Eq. (\ref{eq:N_sources_Fourier_field}) becomes
\begin{align}
     \tilde{E}(k)&=\sqrt{2}\sigma A e^{-k^2\sigma^2}\left(\sum_{j=1}^{N/2}e^{-ikj\delta }e^{-ij\phi}+\sum_{j=1}^{N/2}e^{ikj\delta }e^{ij\phi}\right)\nonumber\\
    &=2\sqrt{2}\sigma A e^{-k^2\sigma^2}\frac{\sin{\left(\frac{N}{4}(k\delta+\phi)\right)}}{\sin{\left(\frac{1}{2}(k\delta+\phi)\right)}}\cos{\left(\frac{1}{4}\left(N+2\right)(k\delta+\phi)\right)}.
\end{align}
Using the same steps to obtain Eq. (\ref{eq:FI_coherent_N_sources}), we obtain the per-photon FI for estimating $L$ as
\begin{align}
     \lim_{L\to0}\frac{ F(L)}{NM} =\frac{1}{16\sigma^2}\frac{\left((N+2)\sin\left(\frac{N\phi}{2}\right)-N\sin\left(\frac{(N+2)\phi}{2}\right)\right)^2}{N^3\sin^4\left(\phi/2\right)}.\label{eq:FI_coherent_N_sources_ramp_phase}
\end{align}
In contrast to the step phase, the phase at which the FI is maximized is now a function of $N$. We have numerically confirmed that in the high-$N$ regime (up to $N=1000$), the maximum FI scales linearly with $N$.\if As shown in Appendix C, SPADE also has the same FI as given by Eq. (\ref{eq:FI_coherent_N_sources_ramp_phase}) for the ramp phase.\fi
\renewcommand\theequation{C\arabic{equation}}
\setcounter{equation}{0}
\subsection*{Appendix C: Phase-sensitive detection of N coherent sources}
\if SPADE has been proposed and shown to saturate the QFI for both the incoherent and coherent two-point case \cite{tsang2016quantumtheoryofsuperresolution,tsangcommentonSaleh}. SPADE has also been used for the time domain problem of estimating the temporal separation between two pulses. For the general case of incoherent objects, SPADE can measure the second order moment of the object in the sub-Rayleigh regime. For partially coherent objects, including N partially coherent sources, Kevin et al. \cite{kevin2021partialcoherencemomentsPRA, Kevinliang2023quantum_N_point_partially_coherent_sources}\ have predicted that a better estimation might be possible, although no optimal scheme was proposed. It seems, though not rigorously proved, that SPADE almost always saturates the QFI at least in the case of single-parameter estimation (second moment) in sub-Rayleigh imaging. However, SPADE is typically `hard' to implement experimentally, and the advantage vanishes quickly in the presence of noise and misalignment such as a misaligned centroid. The quantitative effect of such perturbations on SPADE performance is an active area of research. \fi 

In this section, we show that the FI to estimate sub-Rayleigh separations achieved with phase-sensitive schemes such as mode sorting is \textit{equal} to the FI obtained by Fourier-plane measurements. Specifically, we show that for the two-point case \cite{tsangcommentonSaleh,kurdzialek2022backtosourcesroleofloss,giacomokaruseichyk2022resolvingmutuallycoherentpointsources} the FI for an \textit{aligned} mode sorter is given by Eq. (\ref{eq:FI_coherent}) and the results for the $N$-point constellations in Appendix B are given by Eqs. (\ref{eq:FI_coherent_N_sources},\ref{eq:FI_coherent_N_sources_ramp_phase}). Typically implemented by projecting the image plane field to an appropriate modal basis \cite{OptExpress_Parity_Sorting2016,wadood2021experimentalparitysorting,Fabre2020_Optica_SPADE_2D_Crosstalk}, modal projections allow both spatial \cite{tsang2016quantumtheoryofsuperresolution,Steinberg2017BeatingRayleighscurse,superresolution_with_heterodyne,OptExpress_Parity_Sorting2016,sanchezsoto_2016_optica} and temporal \cite{SanchezSoto_Silberhorn2021effectsofcoherenceontemporalresolution} super-resolution of incoherent and coherent sources \cite{SalehResurgencePaper,tsangcommentonSaleh,SalehreplytoTsang,Kevin2021coherenceOptica,wadood2021experimentalparitysorting,hradil2021exploringtheultimatelimits,SanchezSoto_fisherinformation_With_Coherence_Optica,kurdzialek2022backtosourcesroleofloss,SanchezSoto_Silberhorn2021effectsofcoherenceontemporalresolution,giacomokaruseichyk2022resolvingmutuallycoherentpointsources,giacomosorelli2025ultimate,tsang2021poisson}, and for arbitrary objects in the sub-Rayleigh regime \cite{zhou2019modern,tsang2019quantum,datta2021sub_rayleigh_characterization,sorelli2021momentbasedsuperresolution,frank2023passivelvovskypassivesuperresolution,Lvovsky_pushkina2021superresolution_PRL,Yiyu2021confocal_mode_sorter,tsang2020semiparametric,tsang2019semiparametric}. Although robust to partial or complete incoherence \cite{tsang2016quantumtheoryofsuperresolution,tsang2021poisson,wadood2021experimentalparitysorting,SalehreplytoTsang}, mode sorting is fragile and it fails in the presence of misalignment or an unknown centroid \cite{tsang2016quantumtheoryofsuperresolution,Michael_Grace_JOSAB_unknown_Centroid,vrehavcek_sanchez_soto_2017multiparameter,hervas2024optimizingsanchezsotoetradeoffsinmultiparametersuperresolution,de2021discrimination_under_misalignment,schodt2023tolerance_to_aberration_and_misalignment,optimal_observables_for_practical_superresolution,PRL2020_superresolution_limits_frm_crosstalk}.\par 
For clarity of presentation, it proves convenient to use bra-ket notation, although our analysis is purely classical. Detection of a given mode $m$ can then be written as a projection $\langle m|E\rangle$, where $|E\rangle$ is the original field. As we have assumed Gaussian PSFs, we can represent $|E\rangle$ as a superposition of shifted coherent states $|\alpha_{i}\rangle$, where $\alpha_{i}=\delta_{i}/2\sigma$. For the sub-Rayleigh case, it is sufficient to consider field projections on the first-order HG mode $|m=1\rangle$ \cite{steinberg2017beating_rayleighscurse}. 

For a knife-edge style phase, the field in Eq. (\ref{eq:N_point_field}) can  be written 
\begin{align}
    |E\rangle&=\sum_{j=-1}^{-N/2}|\alpha_{j}\rangle+e^{i\phi}\sum_{j=1}^{N/2}|\alpha_{j}\rangle \nonumber\\
  &=  \sum_{j=1}^{N/2}\left(|\alpha_{-j}\rangle+e^{i\phi}|\alpha_{j}\rangle \right).\label{eq:SPADE_field_knife_edge}
\end{align}
where we have assumed that each source emits one photon, i.e. we set $A_j=(2\pi\sigma^2)^{-1/4}$. Using $ \alpha_{-j}=-\alpha_{j}$ and $\langle n|\alpha\rangle=e^{-|\alpha|^2/2}\alpha^n/\sqrt{n!}$, projection onto the first-order mode $|1\rangle$ gives
\begin{align}
    \langle1|E\rangle&=\sum_{j=1}^{N/2}\left(\langle 1|\alpha_{-j}\rangle+e^{i\phi}\langle 1|\alpha_{j}\rangle \right)\nonumber\\
    &=(e^{i\phi}-1)\sum_{j=1}^{N/2}\left(e^{-|\alpha_{j}|^2/2}\alpha_{j}\right).
\end{align}
The corresponding intensity is 
\begin{align}
    P(\delta)&=|\langle1|E\rangle|^2=2\left(1-\cos{\left(\phi\right)}\right)\sum_{j,l=1}^{N/2}\left(e^{-(|\alpha_{j}|^2+|\alpha_{l}|^2)/2}\alpha_{j}\alpha_{l}\right)\nonumber\\
    &=2\left(1-\cos{\left(\phi\right)}\right)\sum_{j,l=1}^{N/2}\left(e^{-(|\delta_{j}|^2+|\delta_{l}|^2)/8\sigma^2}\frac{\delta_{j}\delta_{l}}{4\sigma^2}\right)
\end{align}
We now assume that $\delta_{j,l}/\sigma\ll1$ for all $j,l$, and the factor of $e^{-(\delta_{j}^2+|\delta_{l}|^2)/8\sigma^2}$ tends to unity in the subdiffraction regime. We then have
\begin{align}
    P(\delta)&=2\left(1-\cos{\left(\phi\right)}\right)\sum_{j,l=1}^{N/2}\left(\frac{\delta_{j}\delta_{l}}{4\sigma^2}\right)\nonumber\\
  &=\left(1-\cos{\left(\phi\right)}\right)\left(\frac{\delta^2}{2\sigma^2}\right)\sum_{j,l=1}^{N/2}ij\nonumber\\
  &= \left(1-\cos{\left(\phi\right)}\right)\left(\frac{\delta^2}{2\sigma^2}\right)\frac{N^2(N+2)^2}{64}\label{eq:SPADE_modal_weight_knife_edge}
\end{align}
The FI to estimate $\delta$ is
\begin{align}
   \lim_{\delta\to0}  F_{PS}(\delta)=\frac{\left(\frac{dP(\delta)}{d\delta}\right)^2}{P(\delta)}=\frac{N^2(N+2)^2}{16\sigma^2}\frac{\left(1-\cos{\left(\phi\right)}\right)}{2},
\end{align}
which is equal to Eq. (\ref{eq:FI_coherent}) for $N=2$ and to Eq. (\ref{eq:FI_coherent_N_sources_delta}) for the $N-$point Fourier-plane measurement, and $PS$ denotes phase-sensitive measurements. Note that for arbitrary $\delta$ and $N=2$, the FI for mode sorting equals Eq. (\ref{eq:FIversusdelta}) \cite{giacomokaruseichyk2022resolvingmutuallycoherentpointsources,kurdzialek2022backtosourcesroleofloss,tsangcommentonSaleh}. \par The FI to estimate $L=N\delta$ is $F_{PS}(\delta)/N^2$, and we obtain
\begin{align}
   \lim_{L\to0} F_{PS}(L)=\frac{(N+2)^2}{16\sigma^2}\frac{\left(1-\cos{\left(\phi\right)}\right)}{2}.
\end{align}
The per-photon FI is obtained by dividing by the total photon number, which is also equal to $N$ since we have assumed each source emits one photon. This gives the per-photon FI
\begin{align}
      \lim_{L\to0}\frac{F_{PS}(L)}{N}= \frac{(N+2)^2}{N}\frac{\left(1-\cos{\left(\phi\right)}\right)}{32\sigma^2},
\end{align}
which is the same as Eq. (\ref{eq:FI_coherent_N_sources}) obtained for a Fourier-plane measurement.\par

For a linear ramp phase, the image-plane field in Eq. (\ref{eq:SPADE_field_knife_edge}) can be rewritten as $\ket{E}=\sum_{j=1}^{N/2}\left(|\alpha_{-j}\rangle e^{-ij\phi}+e^{ij\phi}|\alpha_{j}\rangle \right)$. Using steps similar in derivation of Eq. (\ref{eq:SPADE_modal_weight_knife_edge}), we obtain the projection 
\begin{align}
    P(\delta)=|\langle1|E\rangle|^2=\frac{\delta^2}{64\sigma^2}\frac{\left((N+2)\sin\left(\frac{N\phi}{2}\right)-N\sin\left(\frac{(N+2)\phi}{2}\right)\right)^2}{\sin^4\left(\phi/2\right)}\label{eq:SPADE_modal_weight_ramp_phase}.
\end{align}
The per-photon FI to estimate $L$ from Eq. (\ref{eq:SPADE_modal_weight_ramp_phase}) is then same as Eq. (\ref{eq:FI_coherent_N_sources_ramp_phase}) obtained from Fourier measurements.


\begin{thebibliography}{10}
\newcommand{\enquote}[1]{``#1''}

\bibitem{abbe1873articlebeitrage}
E.~Abbe, \enquote{Beitr{\"a}ge zur theorie des mikroskops und der mikroskopischen wahrnehmung,} {\protect\JournalTitle{Archiv f{\"u}r mikroskopische Anatomie}} \textbf{9}, 413--468 (1873).

\bibitem{rayleigh1879xxxi}
L.~Rayleigh, \enquote{Xxxi. {I}nvestigations in {o}ptics, with special reference to the spectroscope,} {\protect\JournalTitle{The London, Edinburgh, and Dublin Philosophical Magazine and Journal of Science}} \textbf{8}, 261--274 (1879).

\bibitem{hellSTED1994breaking}
S.~W. Hell and J.~Wichmann, \enquote{Breaking the diffraction resolution limit by stimulated emission: stimulated-emission-depletion fluorescence microscopy,} {\protect\JournalTitle{Optics letters}} \textbf{19}, 780--782 (1994).

\bibitem{pavani2009doublehelix}
S.~R.~P. Pavani, M.~A. Thompson, J.~S. Biteen, S.~J. Lord, N.~Liu, R.~J. Twieg, and R.~Piestun, \enquote{Three-dimensional, single-molecule fluorescence imaging beyond the diffraction limit by using a double-helix point spread function,} {\protect\JournalTitle{Proceedings of the National Academy of Sciences}} \textbf{106}, 2995--2999 (2009).

\bibitem{rust2006subSTORM}
M.~J. Rust, M.~Bates, and X.~Zhuang, \enquote{Sub-diffraction-limit imaging by stochastic optical reconstruction microscopy (storm),} {\protect\JournalTitle{Nature methods}} \textbf{3}, 793--796 (2006).

\bibitem{betzig1991breakingnearfieldmicroscopy}
E.~Betzig, J.~K. Trautman, T.~Harris, J.~Weiner, and R.~Kostelak, \enquote{Breaking the diffraction barrier: optical microscopy on a nanometric scale,} {\protect\JournalTitle{Science}} \textbf{251}, 1468--1470 (1991).

\bibitem{sanchez1999nearfieldnovotny}
E.~J. S{\'a}nchez, L.~Novotny, and X.~S. Xie, \enquote{Near-field fluorescence microscopy based on two-photon excitation with metal tips,} {\protect\JournalTitle{Physical Review Letters}} \textbf{82}, 4014 (1999).

\bibitem{he2021ultrahigh}
M.~He, G.~R. Iyer, S.~Aarav, S.~S. Sunku, A.~J. Giles, T.~G. Folland, N.~Sharac, X.~Sun, J.~Matson, S.~Liu \emph{et~al.}, \enquote{Ultrahigh-resolution, label-free hyperlens imaging in the mid-ir,} {\protect\JournalTitle{Nano letters}} \textbf{21}, 7921--7928 (2021).

\bibitem{Jason2013nonlinear_Abbe_theory}
C.~Barsi and J.~W. Fleischer, \enquote{Nonlinear {A}bbe theory,} {\protect\JournalTitle{Nature Photonics}} \textbf{7}, 639--643 (2013).

\bibitem{tsang2016quantumtheoryofsuperresolution}
M.~Tsang, R.~Nair, and X.-M. Lu, \enquote{Quantum theory of superresolution for two incoherent optical point sources,} {\protect\JournalTitle{Physical Review X}} \textbf{6}, 031033 (2016).

\bibitem{GburReview}
G.~Gbur, \enquote{Using superoscillations for superresolved imaging and subwavelength focusing,} {\protect\JournalTitle{Nanophotonics}} \textbf{8}, 205 -- 225 (2018).

\bibitem{dertinger2009SOFI}
T.~Dertinger, R.~Colyer, G.~Iyer, S.~Weiss, and J.~Enderlein, \enquote{Fast, background-free, 3d super-resolution optical fluctuation imaging (sofi),} {\protect\JournalTitle{Proceedings of the National Academy of Sciences}} \textbf{106}, 22287--22292 (2009).

\bibitem{hugodefienne2022pixelsuperresolution}
H.~Defienne, P.~Cameron, B.~Ndagano, A.~Lyons, M.~Reichert, J.~Zhao, A.~R. Harvey, E.~Charbon, J.~W. Fleischer, and D.~Faccio, \enquote{Pixel super-resolution with spatially entangled photons,} {\protect\JournalTitle{Nature communications}} \textbf{13}, 3566 (2022).

\bibitem{wadood2021experimentalparitysorting}
S.~A. Wadood, K.~Liang, Y.~Zhou, J.~Yang, M.~Alonso, X.-F. Qian, S.~H. Rafsanjani, A.~N. Jordan, R.~W. Boyd \emph{et~al.}, \enquote{Experimental demonstration of superresolution of partially coherent light sources using parity sorting,} {\protect\JournalTitle{Optics express}} \textbf{29}, 22034--22043 (2021).

\bibitem{giacomokaruseichyk2022resolvingmutuallycoherentpointsources}
I.~Karuseichyk, G.~Sorelli, M.~Walschaers, N.~Treps, and M.~Gessner, \enquote{Resolving mutually-coherent point sources of light with arbitrary statistics,} {\protect\JournalTitle{Physical Review Research}} \textbf{4}, 043010 (2022).

\bibitem{steinberg2017beating_rayleighscurse}
W.-K. Tham, H.~Ferretti, and A.~M. Steinberg, \enquote{Beating {R}ayleigh’s curse by imaging using phase information,} {\protect\JournalTitle{Physical review letters}} \textbf{118}, 070801 (2017).

\bibitem{vrehavcek_sanchez_soto_2017multiparameter}
J.~{\v{R}}eha{\v{c}}ek, Z.~Hradil, B.~Stoklasa, M.~Pa{\'u}r, J.~Grover, A.~Krzic, and L.~S{\'a}nchez-Soto, \enquote{Multiparameter quantum metrology of incoherent point sources: towards realistic superresolution,} {\protect\JournalTitle{Physical Review A}} \textbf{96}, 062107 (2017).

\bibitem{kevin2021partialcoherencemomentsPRA}
K.~Liang, S.~A. Wadood, and A.~Vamivakas, \enquote{Coherence effects on estimating general sub-{R}ayleigh object distribution moments,} {\protect\JournalTitle{Physical Review A}} \textbf{104}, 022220 (2021).

\bibitem{goodman_Fourier_Optics_book}
J.~W. Goodman, \emph{Introduction to {F}ourier {O}ptics} (Roberts and Company Publishers, 2005).

\bibitem{Trepshsu2004optimaldisplacementofsingleGaussian}
M.~T. Hsu, V.~Delaubert, P.~K. Lam, and W.~P. Bowen, \enquote{Optimal optical measurement of small displacements,} {\protect\JournalTitle{Journal of Optics B: Quantum and Semiclassical Optics}} \textbf{6}, 495 (2004).

\bibitem{anthonyvella_alonso_2020MLE_fisher_information_tutorial}
A.~Vella and M.~A. Alonso, \enquote{Maximum likelihood estimation in the context of an optical measurement,} in \emph{Progress in Optics,}  vol.~65 (Elsevier, 2020), pp. 231--311.

\bibitem{Sanchez_Soto_Tempering_Rayleighs_Curse_2018Optica}
M.~Pa{\'u}r, B.~Stoklasa, J.~Grover, A.~Krzic, L.~L. S{\'a}nchez-Soto, Z.~Hradil, and J.~{\v{R}}eh{\'a}{\v{c}}ek, \enquote{Tempering rayleigh’s curse with {PSF} shaping,} {\protect\JournalTitle{Optica}} \textbf{5}, 1177--1180 (2018).

\bibitem{kurdzialek2022backtosourcesroleofloss}
S.~Kurdzialek, \enquote{Back to sources--the role of losses and coherence in super-resolution imaging revisited,} {\protect\JournalTitle{Quantum}} \textbf{6}, 697 (2022).

\bibitem{goodman2015statisticaloptics}
J.~W. Goodman, \emph{Statistical {O}ptics} (John Wiley \& Sons, 2015).

\bibitem{PRL_lupo2020quantumlimitstoincoherentimagingachievedbyinterferometry}
C.~Lupo, Z.~Huang, and P.~Kok, \enquote{Quantum limits to incoherent imaging are achieved by linear interferometry,} {\protect\JournalTitle{Physical Review Letters}} \textbf{124}, 080503 (2020).

\bibitem{Kevinliang2023quantum_N_point_partially_coherent_sources}
K.~Liang, S.~A. Wadood, and A.~N. Vamivakas, \enquote{Quantum fisher information for estimating n partially coherent point sources,} {\protect\JournalTitle{Optics Express}} \textbf{31}, 2726--2743 (2023).

\bibitem{schelkunoff1943mathematicaltheoryoflineararrays}
S.~A. Schelkunoff, \enquote{A mathematical theory of linear arrays,} {\protect\JournalTitle{The Bell System Technical Journal}} \textbf{22}, 80--107 (1943).

\bibitem{bornandwolf2013principlesofoptics}
M.~Born and E.~Wolf, \emph{Principles of optics: electromagnetic theory of propagation, interference and diffraction of light} (Elsevier, 2013).

\bibitem{nikolov2021metaformnickvamivakasmetaformdesign}
D.~K. Nikolov, A.~Bauer, F.~Cheng, H.~Kato, A.~N. Vamivakas, and J.~P. Rolland, \enquote{Metaform optics: Bridging nanophotonics and freeform optics,} {\protect\JournalTitle{Science Advances}} \textbf{7}, eabe5112 (2021).

\bibitem{mandelandwolf1995book}
L.~Mandel and E.~Wolf, \emph{Optical {C}oherence and {Q}uantum {O}ptics} (Cambridge university press, 1995).

\bibitem{bernien2017probingLukin51atomquantumsimulator}
H.~Bernien, S.~Schwartz, A.~Keesling, H.~Levine, A.~Omran, H.~Pichler, S.~Choi, A.~S. Zibrov, M.~Endres, M.~Greiner \emph{et~al.}, \enquote{Probing many-body dynamics on a 51-atom quantum simulator,} {\protect\JournalTitle{Nature}} \textbf{551}, 579--584 (2017).

\bibitem{masson2020manyOrozcomanybodysignaturesinatomicchains}
S.~J. Masson, I.~Ferrier-Barbut, L.~A. Orozco, A.~Browaeys, and A.~Asenjo-Garcia, \enquote{Many-body signatures of collective decay in atomic chains,} {\protect\JournalTitle{Physical review letters}} \textbf{125}, 263601 (2020).

\bibitem{changKimble2012cavityQEDwithatomicmirrors}
D.~E. Chang, L.~Jiang, A.~Gorshkov, and H.~Kimble, \enquote{Cavity qed with atomic mirrors,} {\protect\JournalTitle{New Journal of Physics}} \textbf{14}, 063003 (2012).

\bibitem{asenjo2019optical}
A.~Asenjo-Garcia, H.~Kimble, and D.~E. Chang, \enquote{Optical waveguiding by atomic entanglement in multilevel atom arrays,} {\protect\JournalTitle{Proceedings of the National Academy of Sciences}} \textbf{116}, 25503--25511 (2019).

\bibitem{rovnynatalie2024nanoscalenvcentersreviewarticle}
J.~Rovny, S.~Gopalakrishnan, A.~C.~B. Jayich, P.~Maletinsky, E.~Demler, and N.~P. de~Leon, \enquote{Nanoscale diamond quantum sensors for many-body physics,} {\protect\JournalTitle{Nature Reviews Physics}} pp. 1--16 (2024).

\bibitem{michaelgrace2021quantumoptimal_obect_classification}
M.~R. Grace and S.~Guha, \enquote{Quantum-optimal object discrimination in sub-diffraction incoherent imaging,} {\protect\JournalTitle{arXiv preprint arXiv:2107.00673}}  (2021).

\bibitem{zhang2024performance_partiallycoherenthypothesistesting}
J.-D. Zhang, K.~Zhang, L.~Hou, and S.~Wang, \enquote{Performance advantage of quantum hypothesis testing for partially coherent optical sources,} {\protect\JournalTitle{Journal of the Optical Society of America B}} \textbf{41}, 2540--2547 (2024).

\bibitem{zhang2025quantum_partiallycoherenthypothesistesting}
J.-D. Zhang, M.-M. Zhang, F.~Jia, C.~Li, and S.~Wang, \enquote{Quantum-optimal hypothesis testing for discriminating partially coherent optical sources,} {\protect\JournalTitle{Physical Review A}} \textbf{111}, 023706 (2025).

\bibitem{fienup1982phaseretrieval}
J.~R. Fienup, \enquote{Phase retrieval algorithms: a comparison,} {\protect\JournalTitle{Applied optics}} \textbf{21}, 2758--2769 (1982).

\bibitem{dresselandrewjordan2014colloquiumweakvalues}
J.~Dressel, M.~Malik, F.~M. Miatto, A.~N. Jordan, and R.~W. Boyd, \enquote{Colloquium: Understanding quantum weak values: Basics and applications,} {\protect\JournalTitle{Reviews of Modern Physics}} \textbf{86}, 307--316 (2014).

\bibitem{sethlloydgiovannetti2011advancesinquantummetrology}
V.~Giovannetti, S.~Lloyd, and L.~Maccone, \enquote{Advances in quantum metrology,} {\protect\JournalTitle{Nature photonics}} \textbf{5}, 222--229 (2011).

\bibitem{tsang2021poisson}
M.~Tsang, \enquote{Poisson quantum information,} {\protect\JournalTitle{Quantum}} \textbf{5}, 527 (2021).

\bibitem{Michael_Grace_JOSAB_unknown_Centroid}
M.~R. Grace, Z.~Dutton, A.~Ashok, and S.~Guha, \enquote{Approaching quantum-limited imaging resolution without prior knowledge of the object location,} {\protect\JournalTitle{JOSA A}} \textbf{37}, 1288--1299 (2020).

\bibitem{hervas2024optimizingsanchezsotoetradeoffsinmultiparametersuperresolution}
J.~Hervas, L.~S{\'a}nchez-Soto, A.~Goldberg, Z.~Hradil, and J.~{\v{R}}eh{\'a}{\v{c}}ek, \enquote{Optimizing measurement tradeoffs in multiparameter spatial superresolution,} {\protect\JournalTitle{Physical Review A}} \textbf{110}, 033716 (2024).

\bibitem{de2021discrimination_under_misalignment}
J.~de~Almeida, J.~Ko{\l}ody{\'n}ski, C.~Hirche, M.~Lewenstein, and M.~Skotiniotis, \enquote{Discrimination and estimation of incoherent sources under misalignment,} {\protect\JournalTitle{Physical Review A}} \textbf{103}, 022406 (2021).

\bibitem{schodt2023tolerance_to_aberration_and_misalignment}
D.~J. Schodt, P.~J. Cutler, F.~E. Becerra, and K.~A. Lidke, \enquote{Tolerance to aberration and misalignment in a two-point-resolving image inversion interferometer,} {\protect\JournalTitle{Optics Express}} \textbf{31}, 16393--16405 (2023).

\bibitem{optimal_observables_for_practical_superresolution}
G.~Sorelli, M.~Gessner, M.~Walschaers, and N.~Treps, \enquote{Optimal observables and estimators for practical superresolution imaging,} {\protect\JournalTitle{Physical Review Letters}} \textbf{127}, 123604 (2021).

\bibitem{PRL2020_superresolution_limits_frm_crosstalk}
M.~Gessner, C.~Fabre, and N.~Treps, \enquote{Superresolution limits from measurement crosstalk,} {\protect\JournalTitle{Physical Review Letters}} \textbf{125}, 100501 (2020).

\bibitem{laurawaller2012phasespacemeasurementandcoherencesynthesis}
L.~Waller, G.~Situ, and J.~W. Fleischer, \enquote{Phase-space measurement and coherence synthesis of optical beams,} {\protect\JournalTitle{Nature Photonics}} \textbf{6}, 474--479 (2012).

\bibitem{tsangcommentonSaleh}
M.~Tsang and R.~Nair, \enquote{Resurgence of {R}ayleigh{'}s curse in the presence of partial coherence: comment,} {\protect\JournalTitle{Optica}} \textbf{6}, 400--401 (2019).

\bibitem{OptExpress_Parity_Sorting2016}
Z.~S. Tang, K.~Durak, and A.~Ling, \enquote{Fault-tolerant and finite-error localization for point emitters within the diffraction limit,} {\protect\JournalTitle{Optics express}} \textbf{24}, 22004--22012 (2016).

\bibitem{Fabre2020_Optica_SPADE_2D_Crosstalk}
P.~Boucher, C.~Fabre, G.~Labroille, and N.~Treps, \enquote{Spatial optical mode demultiplexing as a practical tool for optimal transverse distance estimation,} {\protect\JournalTitle{Optica}} \textbf{7}, 1621--1626 (2020).

\bibitem{Steinberg2017BeatingRayleighscurse}
W.-K. Tham, H.~Ferretti, and A.~M. Steinberg, \enquote{Beating {R}ayleigh’s curse by imaging using phase information,} {\protect\JournalTitle{Physical review letters}} \textbf{118}, 070801 (2017).

\bibitem{superresolution_with_heterodyne}
F.~Yang, A.~Tashchilina, E.~S. Moiseev, C.~Simon, and A.~I. Lvovsky, \enquote{Far-field linear optical superresolution via heterodyne detection in a higher-order local oscillator mode,} {\protect\JournalTitle{Optica}} \textbf{3}, 1148--1152 (2016).

\bibitem{sanchezsoto_2016_optica}
M.~Pa{\'u}r, B.~Stoklasa, Z.~Hradil, L.~L. S{\'a}nchez-Soto, and J.~Rehacek, \enquote{Achieving the ultimate optical resolution,} {\protect\JournalTitle{Optica}} \textbf{3}, 1144--1147 (2016).

\bibitem{SanchezSoto_Silberhorn2021effectsofcoherenceontemporalresolution}
S.~De, J.~Gil-Lopez, B.~Brecht, C.~Silberhorn, L.~L. S{\'a}nchez-Soto, Z.~Hradil, and J.~{\v{R}}eh{\'a}{\v{c}}ek, \enquote{Effects of coherence on temporal resolution,} {\protect\JournalTitle{Physical Review Research}} \textbf{3}, 033082 (2021).

\bibitem{SalehResurgencePaper}
W.~Larson and B.~E. Saleh, \enquote{Resurgence of {R}ayleigh{'}s curse in the presence of partial coherence,} {\protect\JournalTitle{Optica}} \textbf{5}, 1382--1389 (2018).

\bibitem{SalehreplytoTsang}
W.~Larson and B.~E. Saleh, \enquote{Resurgence of {R}ayleigh’s curse in the presence of partial coherence: reply,} {\protect\JournalTitle{Optica}} \textbf{6}, 402--403 (2019).

\bibitem{Kevin2021coherenceOptica}
K.~Liang, S.~A. Wadood, and A.~N. Vamivakas, \enquote{Coherence effects on estimating two-point separation,} {\protect\JournalTitle{Optica}} \textbf{8}, 243--248 (2021).

\bibitem{hradil2021exploringtheultimatelimits}
Z.~Hradil, D.~Koutn{\`y}, and J.~{\v{R}}eh{\'a}{\v{c}}ek, \enquote{Exploring the ultimate limits: super-resolution enhanced by partial coherence,} {\protect\JournalTitle{Optics Letters}} \textbf{46}, 1728--1731 (2021).

\bibitem{SanchezSoto_fisherinformation_With_Coherence_Optica}
Z.~Hradil, J.~{\v{R}}eh{\'a}{\v{c}}ek, L.~S{\'a}nchez-Soto, and B.-G. Englert, \enquote{Quantum {F}isher information with coherence,} {\protect\JournalTitle{Optica}} \textbf{6}, 1437--1440 (2019).

\bibitem{giacomosorelli2025ultimate}
G.~Sorelli, M.~Gessner, and F.~Schlawin, \enquote{Ultimate resolution limits in coherent anti-stokes raman scattering imaging,} {\protect\JournalTitle{arXiv preprint arXiv:2508.01026}}  (2025).

\bibitem{zhou2019modern}
S.~Zhou and L.~Jiang, \enquote{Modern description of {R}ayleigh's criterion,} {\protect\JournalTitle{Physical Review A}} \textbf{99}, 013808 (2019).

\bibitem{tsang2019quantum}
M.~Tsang, \enquote{Quantum limit to subdiffraction incoherent optical imaging,} {\protect\JournalTitle{Physical Review A}} \textbf{99}, 012305 (2019).

\bibitem{datta2021sub_rayleigh_characterization}
C.~Datta, Y.~L. Len, K.~{\L}ukanowski, K.~Banaszek, and M.~Jarzyna, \enquote{Sub-rayleigh characterization of a binary source by spatially demultiplexed coherent detection,} {\protect\JournalTitle{Optics Express}} \textbf{29}, 35592--35601 (2021).

\bibitem{sorelli2021momentbasedsuperresolution}
G.~Sorelli, M.~Gessner, M.~Walschaers, and N.~Treps, \enquote{Moment-based superresolution: Formalism and applications,} {\protect\JournalTitle{Physical Review A}} \textbf{104}, 033515 (2021).

\bibitem{frank2023passivelvovskypassivesuperresolution}
J.~Frank, A.~Duplinskiy, K.~Bearne, and A.~Lvovsky, \enquote{Passive superresolution imaging of incoherent objects,} {\protect\JournalTitle{Optica}} \textbf{10}, 1147--1152 (2023).

\bibitem{Lvovsky_pushkina2021superresolution_PRL}
A.~Pushkina, G.~Maltese, J.~Costa-Filho, P.~Patel, and A.~Lvovsky, \enquote{Superresolution linear optical imaging in the far field,} {\protect\JournalTitle{Physical review letters}} \textbf{127}, 253602 (2021).

\bibitem{Yiyu2021confocal_mode_sorter}
K.~K. Bearne, Y.~Zhou, B.~Braverman, J.~Yang, S.~Wadood, A.~N. Jordan, A.~Vamivakas, Z.~Shi, and R.~W. Boyd, \enquote{Confocal super-resolution microscopy based on a spatial mode sorter,} {\protect\JournalTitle{Optics Express}} \textbf{29}, 11784--11792 (2021).

\bibitem{tsang2020semiparametric}
M.~Tsang, \enquote{Quantum limit to subdiffraction incoherent optical imaging. ii. a parametric-submodel approach,} {\protect\JournalTitle{Physical Review A}} \textbf{104}, 052411 (2021).

\bibitem{tsang2019semiparametric}
M.~Tsang, \enquote{Semiparametric estimation for incoherent optical imaging,} {\protect\JournalTitle{Physical Review Research}} \textbf{1}, 033006 (2019).

\end{thebibliography}

\end{document}